\documentclass{article}
\usepackage{amsmath}

\def\annexes{\medskip\setcounter{section}{0}\def\thesection{\Alph{section}}
\numberwithin{equation}{section}}

\def\qa{q_\alpha}
\def\xa{x_\alpha}
\def\qb{q_\beta}
\def\xb{x_\beta}
\def\a{\alpha}
\def\b{\beta}
\def\r#1{(\ref{#1})}
\def\bin#1#2{\left(#1\atop#2\right)}

\title{Ground state energy  of a non-integer number of particles with $\delta$
attractive interactions}
\author{\'Eric Brunet\footnote{\texttt{Eric.Brunet@physique.ens.fr}}
 \ and Bernard Derrida\footnote{\texttt{Bernard.Derrida@physique.ens.fr}}\\[2mm]
 \small\em Laboratoire de Physique Statistique, \'Ecole Normale
 Sup\'erieure,\\[-1mm]
 \small\em 24, rue Lhomond, 75231	Paris C\'edex 05, France\\[2mm]
 {Physica A \textbf{279} (2000) 395--407}}

\begin{document}
\maketitle
\begin{abstract}
We show how to define and calculate the ground state energy of a system
of quantum particles with $\delta$ attractive interactions when the number
of particles $n$ is non-integer.  The question is relevant to obtain the
probability distribution of the free energy of a directed polymer in a
random medium.  When one expands the ground state energy in powers of the
interaction, all the coefficients of the perturbation series are
polynomials in $n$, allowing to define the perturbation theory for
non-integer $n$. We develop a procedure to calculate all the cumulants of
the free energy of the directed polymer and we give explicit,
although complicated, expressions of the first three cumulants.

\smallskip
\noindent PACS numbers: 05.40.+j, 02.50.-r, 82.20.-w.
\end{abstract}

\noindent{\em 
It is our great pleasure to dedicate this work to our friend 
Joel~L.~Lebowitz on the occasion of his 70th birthday.}

\section{Introduction}
We consider a system of $n$ identical quantum particles on a ring of
size~$L$ with~$\delta$ attractive interactions. If we
call $\xa$ (for $1\le\alpha\le n$) the positions of the particles, the
Hamiltonian of this system is
\begin{equation}
{\cal H}=-{1\over2}\sum_\alpha {\partial^2\over\partial\xa^2}
		-\gamma \sum_{\alpha<\beta}
\delta(\xa-\xb),
\label{hamiltonian}
\end{equation}
where $\gamma$ is the strength of the attractive ($\gamma\ge0$)
interactions.  The main goal of the present work is to define and to
calculate the ground state energy $E_0(n,L,\gamma)$ of (\ref{hamiltonian})
when $n$ is not an integer (especially when $n$ is small).

This system of particles in one dimension with $\delta$ interactions has a
long history in the theory of exactly soluble
models\cite{LiebLiniger.63,YangYang.69,Jimboetall.80,Thacker.81,Gaudin.83}.
It was first introduced to describe a Bose gas by~Lieb and~Liniger who
calculated by Bethe ansatz  the ground state energy and the excitations for
repulsive interactions (that is for negative $\gamma$) in the
thermodynamic limit ($n$ and $L$ go to infinity keeping $n/L$
constant)\cite{LiebLiniger.63}. 

The problem arose also  in the theory of disordered systems: the
calculation of the free energy of a directed polymer in a random medium
in~$1+1$ dimensions by the replica method
reduces\cite{Kardar.87,MezardParisi.91,BouchaudOrland.90,Goldschmidt.93,GarelOrland.97,SaakianNieuwenhuizen.97}
to finding the ground state energy of~\r{hamiltonian}: if  $Z(x,t)$ is
the partition function of a directed polymer joining the points $(0,0)$
and $(x,t)$ on a cylinder with periodic boundary conditions ($x+L \equiv
x$) 
\begin{equation}
Z(x,t) = \int_{(0,0)}^{(x,t)} {\cal D} y(s) \exp \left( - \int_0^t ds
\left[ {1 \over 2  } \left( d y(s) \over ds \right)^2 + \eta(y(s),s)
\right] \right) ,
\label{path}
\end{equation}
where the  random medium is characterised  by a Gaussian white noise
$\eta(y,t)$ 
\begin{equation}
\big\langle \eta(y,t)\, \eta(y',t') \big\rangle = \gamma  \, \delta( y-y')
\, \delta(t-t'), \label{noise}
\end{equation} 
then the integer moments of $Z(x,t)$ are given for large $t$
by\cite{Halpin-HealyZhang.95} 
\begin{equation}
\lim_{t \to \infty} {1 \over t} \ln \left[ \langle Z^n(x,t)\rangle \over
\langle Z(x,t)\rangle^n  \right] = {- E_0(n,L,\gamma)},
\label{relZE}
\end{equation}
where $\langle\ \rangle$ denotes the average over the random medium and $E_0(n,L,\gamma)$ is the ground state energy
of~\r{hamiltonian}.  The knowledge of $E_0(n,L,\gamma)$ determines 
for large $t$ the whole distribution of $\ln  Z(x,t)$.  For
example, the variance of $\ln  Z(x,t)$ is
\begin{equation}
\lim_{t \to \infty} {\langle\ln^2 Z(x,t)\rangle-\langle
\ln  Z(x,t)\rangle^2\over t}= - \ \left.{\partial^2E_0(n,L,\gamma)\over\partial
n^2}\right|_{n=0}.
\end{equation}

Of course, to obtain this variance or other characteristics of the
distribution of $\ln Z(x,t)$, one should be able to define and to calculate
the ground state energy of~\r{hamiltonian} not only for integer~$n$, but
for \emph{any value of~$n$}\cite{Krug.97,Halpin-Healy.91,KrugMeakinHalpin-Healy.92,KimMooreBray.91,Zhang.89}.
Moreover, because of (\ref{noise}), the interactions in~(\ref{hamiltonian})
must be attractive ($\gamma  \geq 0$);  So in contrast to  the Bose gas
initially studied\cite{LiebLiniger.63}, the interactions are attractive and
the interesting limit is no longer the thermodynamic limit $n\to\infty$ but
rather the limit $n\to0$. 

For integer $n$ and $L=\infty$, the $n$ particles form a bound state at
energy\cite{Thacker.81,Kardar.87}
\begin{equation}
E_0(n,\infty,\gamma)=  -\gamma^2 { n ( n^2-1) \over 24}.
\label{kardar1}
\end{equation}
Using this formula for non-integer $n$ helped to  understand several
properties\cite{Halpin-HealyZhang.95} of the distribution of $\ln Z(t)$
when $L$ is infinite.  There are however a number of  difficulties  with
(\ref{kardar1}) for non-integer $n$, in particular a problem of convexity:
$d^2 \ln \langle Z^n \rangle /  d n^2$ should be positive for all $n$, and
so (\ref{relZE}, \ref{kardar1}) cannot be valid at least for negative $n$.
We believe that these difficulties are due to the exchange of limit $t \to
\infty$ and $L \to \infty$ and this is why we try in the present work to
determine $E_0(n,L,\gamma)$ for finite $L$.

The paper is organised as follows: in section~\ref{betheansatz}, we recall
the Bethe ansatz equations which give the ground state energy
of~\r{hamiltonian} for an  (integer) number  $n$ of particles and
we write the integral equation (\ref{eqB})  which
is a way of solving  the coupled non-linear equations of
the Bethe ansatz.  The main advantage of this integral equation
 is that both the  strength~$\gamma$ of the interactions and the
number $n$ of particles appear as continuous parameters. In
section~\ref{smallc} we solve (\ref{eqB}) perturbatively in $c$ (where $c=
\gamma L /2$) for arbitrary $n$. We notice that the coefficients in the
small $c$ expansion of  $E_0(n,L,\gamma)$  are all polynomials in $n$, thus
allowing to define the expansion even for non-integer $n$.  In
section~\ref{smalln}, we show how to generate a small $n$ expansion of the
solution of (\ref{eqB}).  We give explicit expressions  up to order $n^3$
of~$E_0(n,L,\gamma)$ and we notice that the coefficients
of the small~$n$ expansion have in general a zero radius of convergence
in~$c$.

\section{The Bethe ansatz equations}
\label{betheansatz}

The Bethe ansatz\cite{Bethe.31} consists in looking in the region $0\le
x_1\le\dots\le x_n\le L$ for a ground state wave function
$\Psi(x_1,\dots,x_n)$ of the form
\begin{equation}
\Psi(x_1,\dots,x_n)=\sum_{P}  A_P \ 
	e^{{2\over L}(q_1 x_{P(1)}+\dots+q_n x_{P(n)})},
\label{defpsi}
\end{equation}
where the sum in~\r{defpsi} runs
over all the permutations $P$ of $\lbrace 1,\dots,n\rbrace$. The value
of $\Psi$ in  other regions  can be deduced from (\ref{defpsi}) by
symmetries.
One can show\cite{Jimboetall.80,Thacker.81,Gaudin.83} that~\r{defpsi} is
an eigenstate of~\r{hamiltonian} at energy 
\begin{equation}
E(n,L,\gamma)=-{2\over L^2}\sum_{1\le\a\le n}\qa^2,
\label{E0}
\end{equation}
if the $\lbrace\qa\rbrace$ are solutions of the $n$~coupled equations
\begin{equation}
e^{2\qa}=\prod_{\b\ne\a}{\qa-\qb+{c}\over\qa-\qb-{c }},
\label{AnsatzSol}
\end{equation}
where 
\begin{equation}
c={\gamma L\over 2}.
\label{cdef}
\end{equation}
(A derivation of (\ref{AnsatzSol})  can be found in \cite{Jimboetall.80}.
Note that  $i k_j$ and $c$ in \cite{Jimboetall.80} become here respectively
${2\over L}q_j$ and $-\gamma$; the $c$ in \cite{Jimboetall.80} and our $c$
defined in \r{cdef} are thus different.)
Moreover, for $\gamma\ne0$, all the $\qa$ are distinct.

There are \emph{a priori} many solutions of~\r{AnsatzSol}.  We look for the
ground state, that is the solution $\lbrace\qa\rbrace$ for which~\r{E0}
is minimal.  When $c=0$, the problem reduces to
$n$ non-interacting particles $\lbrace\qa\rbrace=\lbrace0\rbrace$ (we have
periodic boundary conditions). Because the ground state solution is not
degenerate,  the solution $\lbrace\qa\rbrace$ of~\r{AnsatzSol} must have
the symmetry  $\lbrace\qa\rbrace = \lbrace- \qa\rbrace$, depend
continuously on $c$, and vanish as $c\to0$.


Let us introduce the following function of $\lbrace \qa \rbrace$:
\begin{equation}
B(u)={1\over n}e^{{c\over4}(u^2-1)}\sum_{\qa} \rho(\qa) e^{\qa(u-1)},
\label{defB}
\end{equation}
where the parameters  $\rho(\qa)$ are defined by
\begin{equation}
\rho(\qa)=\prod_{\qb\ne\qa}{\qa-\qb+c\over \qa-\qb}.
\label{defrho}
\end{equation}
The function $B(u)$ is a rather complicated (but easier to manipulate than
the~$\qa$) symmetric function of the ground state solution
$\lbrace\qa\rbrace$ of (\ref{AnsatzSol}).  As shown  in the appendix, it
satisfies the integral equation
\begin{align}
B(1+u)-B(1-u)
 =nc\int_0^u e^{-{c\over2}(v^2-uv)}B(1-v)B(1+u-v)\,dv,
\label{eqB}
\end{align}
and the following two conditions
\begin{align}
B(1)&=1,
\label{B(1)} \\
B(u)&=B(-u).
\label{paritybis}
\end{align}
Moreover,  the energy (\ref{E0})  can be deduced  from of $B(u)$ by 
\begin{equation}
\label{relEB}
E_0(n,L,\gamma)  = {2 \over L^2} \left[{n^3c^2\over6}+{n c^2\over12}+{n c\over2} - n B''(1) \right].
\end{equation}

The  derivation of (\ref{eqB}, \ref{B(1)}, \ref{paritybis}, \ref{relEB}) is
given in the appendix.  How  these relations lead to  small $c$ or small
$n$ expansions is explained in sections~\ref{smallc} and~\ref{smalln}.

\section{Expansion in powers of $c$}
\label{smallc}

One could try to solve the equations (\ref{AnsatzSol}) perturbatively in
$c$ but the approach turns out to be quickly
complicated\cite{BrunetDerrida2.00}. Instead we are going to see that  the
integral equation (\ref{eqB}) is very convenient to obtain
$E_0(n,L,\gamma)$ for small~$c$.

It is known\cite{Gaudin.83} (and easy to check from (\ref{AnsatzSol}))
that the $\qa$ scale like $\sqrt{c}$ for small~$c$.
Therefore each coefficient $B_i(u)$ of the small $c$ expansion of $B(u)$
defined by (\ref{defB}) is a polynomial in $u$.
\begin{equation}
B(u)=B_0(u)+c B_1(u) +c^2 B_2(u)+\dots
\end{equation}
 Moreover, conditions~\r{B(1)}
and~\r{paritybis} impose that all the $B_i(u)$ are even,
that  $B_0(1)=1$ and $B_i(1)=0$ for  any $i \geq 1 $.

At zero-th order in $c$, we find, using~\r{eqB}:
\begin{equation}
B_0(1+u)-B_0(1-u)=0.
\end{equation}
Thus $B_0(u)$ and $B_0(1+u)$ are both even functions of $u$. As
$B_0(u)$ is a polynomial and  $B_0(1)=1$,   the only solution is
\begin{equation}
B_0(u)=1.
\end{equation}
We put this back into~\r{eqB} and get at first order in
$c$
\begin{equation}
B_1(1+u)-B_1(1-u)= n u.
\end{equation}
Again, using  the fact  that $B_1(u)$ is an even polynomial
such that $B_1(1)=0$,  the only  possible solution is:
\begin{equation}
B_1(u)={n\over4}(u^2-1).
\end{equation}
It is easy to see from (\ref{eqB}) that at any order in $c$, we  have to
solve 
\begin{align}
\label{genB}
B_i(1+u)-B_i(1-u)=
\hbox{``some polynomial odd in $u$''},
\end{align}
and that there is a unique even polynomial solution satisfying $B_i(1)=0$.
One can  generate as many $B_i(u)$ as needed to obtain
$B(u)$ up to any desired order in $c$.
\begin{align}
\label{Bexpn}
B(u)= 1 + { cn(-1+u^2) \over 4} +
   {c^2 n ( 1 + 2n )( -1 + u^2)^2\over 96}
+O(c^3).
\end{align}
 Relation~\r{relEB} then
gives the energy. Up to the fourth order in $c$, we find
\begin{align}
\label{energie}
{L^2\over2} E_0(n,L,\gamma) = -n(n-1)\Bigg(&
{c\over2}+{c^2\over12}+{n\over180}c^3+
\left({n^2\over1512}-{n\over1260}\right)c^4+...\Bigg).
\end{align}

\section{Solution for small $n$}
\label{smalln}

It is clear from section~\ref{smallc} that if we stop the small~$c$
expansion of $B(u)$ at a given order, $B(u)$ and $E_0(n,L,\gamma)$ are
polynomials in $n$.  This allows to define the small $c$ expansion of
$B(u)$ or of $E_0(n,L,\gamma)$ for an arbitrary value of $n$.

Moreover, we can write a small~$n$ expansion of $B(u)$ by collecting all
the terms proportional to $n^k$ in the small $c$ expansion of $B(u)$ and
calling the series obtained $b_k(u)$. Then,
\begin{equation}
\label{nBexpn}
B(u)=1 + n b_1(u) +n^2 b_2(u)+\dots
\end{equation}
Conditions (\ref{B(1)}, \ref{paritybis}) impose that $b_k(u)=b_k(-u)$
and $b_k(1)=0$ for all $k \geq 1$.

We are now going to describe a procedure which leads to a recursion on the
$b_k(u)$ and allows  to write them not only as power series
in $c$ but as explicit functions of $c$ and $u$.
If we insert (\ref{nBexpn}) into (\ref{eqB}), we get
at first order in $n$
\begin{equation}
b_1(1+u)-b_1(1-u)=c\int_0^u e^{-{c\over2}(v^2-uv)}\,dv.
\label{eqb1}
\end{equation}
It can be checked that a solution of (\ref{eqb1})  compatible with the
conditions $b_1(u)=b_1(-u)$ and $b_1(1)=0$ is
\begin{equation}
b_1(u)=\sqrt c\int_0^{+\infty}\kern-1.3em d\lambda\, 
{\cosh{\lambda u\sqrt c\over2}
-\cosh{\lambda\sqrt c\over2}\over\sinh{\lambda\sqrt
c\over2}}e^{-{\lambda^2\over2}}.
\label{b1}
\end{equation}
There are however other solutions to the difference equation~(\ref{eqb1}):
one could add to (\ref{b1}) an arbitrary function $F(u,c)$ even and
periodic in $u$ of period 2 and vanishing at  $u=1$.  If we require that
each term in the small $c$ expansion of $b_1(u)$ is polynomial in $u$ (as
justified in section~\ref{smallc}), we see that \emph{all the terms of the
small $c$ expansion of $F(u,c)$ must be identically zero}. For example
$F(u,c)=\exp(-c^{-1/4}) (\cos (\pi u)+1)$ is an acceptable function.  This
already shows that $b_1(u)$ given by (\ref{b1})  has indeed for small $c$
expansion the series obtained by collecting all the terms proportional to
$n$ in the small~$c$ expansion of section~\ref{smallc}.

If the solution  $b_1(u)$ (\ref{b1}) had a non-zero radius of convergence
in $c$, it would be natural to choose the only $b_1(u)$ which is analytic
in~$c$ at~$c=0$ by taking $F(u,c)=0$.  Unfortunately, this is not the case:
by making the change of variable $\lambda^2=\mu$, expression~\r{b1}
appears as the Borel sum of a divergent series\cite{Candelpergher.97}.

We found  no conclusive reasons why (\ref{b1}) is the  solution of
(\ref{eqb1}) we should select. However, one can  notice  that, when $n$ is
an integer, $B(u)$ is analytic in $u$ and goes to zero when $u\to \pm i
\infty$  (see~\r{defB}). Here, the $b_1(u)$ given by (\ref{b1}) grows like
$\ln |u|$  when $u\to \pm i \infty$.   Adding a non-zero periodic
$F(u,c)$ would either lead to an exponential growth in the imaginary
direction or introduce singularities in the complex $u$ plane. So
(\ref{b1}) is the  solution of (\ref{eqb1}), analytic in the whole $u$
plane, which has the slowest growth in the imaginary direction.

The same difficulty of selecting the right solution  appears at every order
in the expansion in powers of $n$.  We are now going to  explain the
procedure we have used to select one solution.  If we insert (\ref{nBexpn})
into (\ref{eqB}), we have to solve  at any order $k$ in the small~$n$
expansion
\begin{equation}
b_k(1+u)-b_k(1-u)=\phi_k(u),
\label{genb}
\end{equation}
where $\phi_k(u)$ is a function odd in $u$ which can be calculated 
from the previous orders
\begin{equation}
\phi_k(u)=
c \sum_{i=0}^{k-1} \int_0^u e^{-{c \over 2} (v^2 - uv)} b_i (1-v) b_{k-i-1}
(1 + u - v) dv.
\label{phik}
\end{equation}
(For consistency, we use $b_0(u)=1$.) It can be checked that a solution to
(\ref{genb}) is
\begin{equation}
b_k(u)=\int_0^{+\infty}\kern-1.3em d\lambda\, 
{\cosh{\lambda u\sqrt c\over2}
-\cosh{\lambda\sqrt c\over2}\over\sinh{\lambda\sqrt
c\over2}}a_k(\lambda),
\label{solgenb}
\end{equation}
where $a_k(\lambda) $ is given by
\begin{equation}
a_k(\lambda)={1\over2i\pi}\int_0^{+\infty}\kern-1.3emdu\,\sin{\lambda u
\over2}\phi_k\left({i u\over\sqrt c}\right).
\label{solgena}
\end{equation}
Indeed, the verification is a simple matter of algebra. (We convinced
ourselves that $b_k(u)\sim \ln^k|u|$ and $\phi_k(u)\sim \ln^{k-1}|u|/u$ as
$u\to\pm i\infty$, and that $a_k(\lambda) \sim \ln^{k-1}|\lambda|$ for
$\lambda \to 0$ and $ a_k(\lambda) \sim \exp(-\lambda^2(k+1)/4k)$ for
$\lambda \to \infty$, so that all the integrals in~(\ref{solgenb},
\ref{solgena}) converge.)

As for $b_1(u)$, one could add to the $b_k(u)$ given by~(\ref{solgenb},
\ref{solgena}) an arbitrary function $F_k(u,c)$, even and periodic in $u$
of period 2  to obtain the general solution of (\ref{genb}).  However, to
be consistent with the small $c$ expansion of section~\ref{smallc}, the
small $c$ expansion of  $F_k(u,c)$   should be identically zero.  Moreover if
we want the solution of (\ref{genb}) to be analytic in the whole $u$ plane
and not to grow too fast when $u \to \pm i\infty$, we must take
$F_k(u,c)=0$.

Expression~\r{b1} for $b_1(u)$ is in fact a particular case of the 
procedure~(\ref{solgenb}, \ref{solgena}); when applied
to~\r{eqb1}, it gives indeed $a_1(\lambda)=\sqrt c\exp(-\lambda^2/2)$.

At second order in the small $n$ expansion, we find for $\lambda>0$
\begin{equation}
\label{a2}
a_2(\lambda)
=c e^{-{\lambda^2\over2}}\left[\int_0^{\lambda}\kern-0.8em
d\mu\,e^{-{\mu^2\over2}}{2\cosh{\lambda\mu\over2}-2\over\tanh{\mu\sqrt
c\over2}}
+\int_\lambda^{+\infty}\kern-1.7em 
d\mu\,e^{-{\mu^2\over2}}{e^{-{\lambda\mu\over2}}-2\over\tanh{\mu\sqrt
c\over2}}\right].
\end{equation}
The expressions of $b_3(u)$ or $a_3(\lambda)$ would be much longer to write
and higher orders even more complicated. Recursion~(\ref{phik}, 
\ref{solgenb}, \ref{solgena}) allows nevertheless to calculate in principle
the whole expansion in powers of~$n$.

Using relation~\r{relEB} and the expressions~(\ref{b1}) and~(\ref{a2}) of
$b_1(u)$ and $a_2(\lambda)$, we find that the energy~$E_0(n,L,\gamma)$ is
given up to order $n^3$:
\begin{align}
\label{Enc}
&{L^2\over2}E_0(n,L,\gamma)=
n\left({c\over2}+{c^2\over12}\right)-
n^2 {c^{3/2}\over4}\int_0^{+\infty}\kern-1.7em
d\lambda{\lambda^2\over\tanh{\lambda\sqrt c\over2}}
e^{-{\lambda^2\over2}}+n^3{c^2\over6}\\
&\qquad -n^3 {c^2\over4}\int_0^{+\infty}\kern-1.7em
d\lambda{\lambda^2e^{-{\lambda^2\over2}}\over\tanh{\lambda\sqrt c\over2}}
\Bigg[\int_0^{\lambda}\kern-0.8em
d\mu\,e^{-{\mu^2\over2}}{2\cosh{\lambda\mu\over2}-2\over\tanh{\mu\sqrt
c\over2}}
+\int_\lambda^{+\infty}\kern-1.7em 
d\mu\,e^{-{\mu^2\over2}}{e^{-{\lambda\mu\over2}}-2\over\tanh{\mu\sqrt
c\over2}}\Bigg]
.\nonumber
\end{align}
As explained in the introduction, this small $n$ expansion of
$E_0(n,L,\gamma)$ gives the cumulants of the free energy in the directed
polymer problem.
Of course, if we expand~\r{Enc} in powers of $c$, we recover~\r{energie}.

\section{Conclusion}

In this work, we have developed a method  allowing to calculate
perturbatively the ground
state energy of~(\ref{hamiltonian}) for a
non-integer number $n$ of particles.  We first generated for
integer $n$ a perturbation series in powers of the interaction~$c$. Each
term of this series is polynomial in $n$, allowing to define
a small~$c$ expansion of the energy for non-integer~$n$.
This series, at least for small~$n$, has in general a zero radius of
convergence, in contrast to integer~$n$ for which the radius of convergence
of the perturbation theory is non-zero\cite{Kato.66}. (For $n=2$, the
closest singularities of $E_0(2,L,\gamma)$ in the $c$ plane lie at $c\simeq
3.30\pm i\,4.12$.)

We believe that the fact that each term in the perturbation theory is
polynomial in $n$ is generic and would be true for an arbitrary pair
interaction and in any dimension. As the link~\r{relZE} to directed
polymers is valid in any dimension, it would be useful and interesting to
try to recover our results by doing a direct perturbation theory of the
Hamiltonian instead of our Bethe ansatz approach (which is limited to $1+1$
dimensions and to a $\delta$ potential) in order  to see whether the
calculations could be extended to higher dimensions.

Our calculation of  the ground state energy for non-integer $n$ is based on
the  integral equation (\ref{eqB}) and the conditions
(\ref{B(1)}, \ref{paritybis}).  When we tried to solve the problem for
small $n$, at each order we had to select a particular solution of a
difference equation.  We did not find a conclusive reason to justify the
solution we selected, apart from  some analyticity properties and growth
criterion in the complex plane of the variable $u$.  It would certainly be
interesting to justify our choice~(\ref{Enc}) by  calculating  the
second and the third cumulants of $\ln Z$ directly (and not only
perturbatively to all orders in $c$).

In our  small $n$ expansion of section~\ref{smalln}, the terms become
quickly very complicated. There is however a regime, which corresponds to
the large $c$ limit of recursion~(\ref{phik}, \ref{solgenb},
\ref{solgena}) where one can handle all orders  in the small~$n$
expansion\cite{BrunetDerrida2.00}.  This allows one to calculate the whole
distribution of $\ln(Z(x,t))/t$ when $t$ is very large and
$(1/t)\ln(Z(x,t)/\langle Z(x,t)\rangle)$ of order~$1/L$.  One can then
recover\cite{BrunetDerrida2.00} the same large deviation function as found
for the asymmetric exclusion
process\cite{DerridaLebowitz.98,DerridaAppert.99,Kim.95,LeeKim.99,Appert.00},
as expected since the directed polymer problem in $1+1$ dimensions and the
asymmetric exclusion process are both representatives of the KPZ
equation\cite{KardarParisiZhang.86,Halpin-HealyZhang.95,BarabasiStanley.95}
in dimension $1$.  This strengthens the conjecture that the solutions to
the difference equations we selected in section~\ref{smalln} give indeed
the  right non-integer moments of the partition function.

From the point of view of the theory of disordered systems, our results
give one of the very few examples for which the distribution of $Z$ can be
calculated exactly. In particular  they could provide a good test of the
replica approach and of other variational
methods\cite{MezardParisi.91,Goldschmidt.93,GarelOrland.97,SaakianNieuwenhuizen.97}.

A simple and interesting phenomenon visible in the present work  (which
is generic of all kinds of disordered systems with Gaussian
disorder) is that the weak disorder expansion (here small $c$ expansion) of
non-integer moments  of the partition function has a zero radius of
convergence whereas integer moments have a non-zero radius of convergence.
This is already visible in the trivial example   of a single Ising spin
$\sigma= \pm 1$ in a random Gaussian field~$h$; the partition function
at temperature~$T$ is $Z=2\cosh(h/T)$, and it is easy to check that all
non-integer moments of the partition function have a zero radius of
convergence in $1/T$.

\medskip

\noindent{\bf Acknowledgements}
We thank Fran\c{c}ois David, Michel Gaudin,  Vincent Pasquier, Leonid
Pastur, Herbert Spohn and Andr{\'e} Voros
for useful discussions.

\annexes

\section{Derivation of (\ref{eqB}, \ref{B(1)}, \ref{paritybis}, \ref{relEB})}

Let us first establish some useful properties of the numbers $\rho(\qa)$
defined by~(\ref{defrho}).  If the $\qa$ are the $n$  roots of the
polynomial $P(X)$ defined as 
\begin{equation}
P(X) = \prod_{\qa} (X - \qa),
\label{defpoly}
\end{equation}
it is easy to see that the $\rho(\qa)$  defined in (\ref{defrho})
satisfy
\begin{equation}
{P(X+c)\over P(X)}=1+c\sum_{\qa} {\rho(\qa)\over X-\qa}.
\label{simpleelement}
\end{equation}
(The two sides have the same poles with the same residues and coincide 
at $X \to \infty$.) Expanding the right hand side
of~\r{simpleelement} for large $X$, we get
\begin{align}
{P(X+c) \over P(X)} = 1 + 
c\sum_{\qa}{\rho(\qa)\over X}\left(1+{\qa\over X}+{\qa^2\over
X^2}\right) + O \left( 1 \over X^4 \right).
\label{rap1}
\end{align}
On the other hand, using (\ref{E0}, \ref{defpoly})  and the
 symmetry   $\lbrace\qa\rbrace = \lbrace - \qa\rbrace$  we have
\begin{equation}
P(X)= X^n  + {L^2 \over 4} E_0(n,L,\gamma) X^{n-2} + O(X^{n-4}),
\end{equation}
so that 
\begin{equation}
{P(X+c) \over P(X)} = 1+ {nc\over X} + {c^2\bin{n}{2}\over
X^2}+{c^3\bin{n}{3}-c E_0(n,L,\gamma) L^2 / 2 \over X^3} + O \left( 1 \over
X^4 \right).
\label{rap2}
\end{equation}
Comparing (\ref{rap1}) and (\ref{rap2}), we get the relations
\begin{align}
\label{prop1}
\sum_{\qa} \rho(\qa)&=n,\\
\sum_{\qa} \qa\rho(\qa)&=c\bin{n}{2}, \label{prop3}\\
\sum_{\qa} \qa^2\rho(\qa)&=c^2\bin{n}{3}- {E_0(n,L,\gamma) L
^2 \over 2}  \label{prop4}
\end{align}
Moreover, by letting $X=\pm\qb-c$ in~\r{simpleelement} 
 we get for any $\qb$ root of $P(X)$
\begin{equation}
\label{prop2}
{1\over c}=
\sum_{\qa}{\rho(\qa)\over\qa-\qb+c}
=\sum_{\qa}{\rho(\qa)\over\qa+\qb+c}.
\end{equation}
Lastly using the symmetry $\lbrace\qa\rbrace = \lbrace - \qa\rbrace$ and
the definition (\ref{defrho}), the Bethe ansatz equations
(\ref{AnsatzSol}) reduce  to 
\begin{equation}
e^{\qa} \rho(-\qa)-e^{-\qa}\rho(\qa)=0.
\label{eqrho}
\end{equation}

From the definition (\ref{defB}) of $B(u)$ and the properties
(\ref{prop1}--\ref{eqrho}), it is straightforward to establish
(\ref{eqB}--\ref{relEB}): the integral equation (\ref{eqB}) is a direct
consequence of (\ref{defB}) and (\ref{prop2}). Properties
(\ref{B(1)}, \ref{paritybis}) follow from (\ref{defB}, \ref{prop1}) and
(\ref{defB}, \ref{eqrho}) respectively. Lastly (\ref{relEB}) is a
consequence of (\ref{defB}, \ref{prop1}--\ref{prop4}).

\end{document}